\documentclass[aip,apl, reprint, floatfix]{revtex4-1}

\usepackage{graphicx}
\usepackage{bm}
\usepackage{natbib}
\usepackage{textcomp}
\usepackage{array}

\usepackage[utf8]{inputenc}
\usepackage[english]{babel}
 
\usepackage[usenames, dvipsnames]{color}

\newcolumntype{C}[1]{>{\centering\arraybackslash}m{#1}}

\makeatletter
\let\origsection\section
\renewcommand\section{\@ifstar{\starsection}{\nostarsection}}

\newcommand\nostarsection[1]{\sectionprelude\origsection{#1}\sectionpostlude}

\newcommand\starsection[1]{\sectionprelude\origsection*{#1}\sectionpostlude}

\let\origsubsection\subsection
\renewcommand\subsection{\@ifstar{\starsubsection}{\nostarsubsection}}

\newcommand\nostarsubsection[1]{\sectionprelude\origsubsection{#1}\sectionpostlude}

\newcommand\starsubsection[1]{\sectionprelude\origsubsection*{#1}\sectionpostlude}

\newcommand\sectionprelude{%
  \vspace{-2mm}
}

\newcommand\sectionpostlude{%
  \vspace{-1mm}
}

\makeatother

\draft 

\begin{document}


\title{Robust Modeling of Acoustic Phonon Transmission in Nanomechanical Structures} 



\author{J. Bartlett}
\email{jdb382@cornell.edu}
\affiliation{Sibley School of Mechanical Engineering, Cornell University, Ithaca, New York 14853}
\author{K. Rostem}
\email{karwan.rostem@nasa.gov}
\affiliation{NASA Goddard Spaceflight Center, 8800 Greenbelt Road, Greenbelt, Maryland 20771}
\author{E. J. Wollack}
\affiliation{NASA Goddard Spaceflight Center, 8800 Greenbelt Road, Greenbelt, Maryland 20771}


\date{\today}

\begin{abstract}
The transmission of acoustic phonons is an important element in the design and performance of nano-mechanical devices operating in the mesoscopic limit. Analytic expressions for the power transmission coefficient, $\mathcal{T}$, exist only in the low-frequency (quasi-static) limit described by thin-plate elastic theory, and for well-defined elastic wave-guiding geometries. We compare two numerical techniques based on finite-element computations to determine the frequency dependence of $\mathcal{T}$ for arbitrary phonon scattering structures. Both methods take into account acoustic mode conversion to acoustic and optical modes. In one case, phase and amplitude of complex-valued reflected waves are determined and related to transmission through a Fresnel equation, while in the other the magnitude of the transmitted mechanical power is directly calculated. The numerical robustness of these methods is demonstrated by considering the transmission across an abrupt junction in a rectangular elastic beam, a well-known problem of considerable importance in mesoscopic device physics. The simulations presented extend the standard results for acoustic phonon transmission at an abrupt junction, and are in good agreement with analytic predictions from thin-plate elastic theory in the long-wavelength limit. More generally, the numerical methods developed provide an effective tool for calculating acoustic mode energy loss in nano-mechanical resonators through mode conversion and heat transfer in arbitrary mesoscopic structures.




\end{abstract}


\maketitle 

\section{Introduction}

The rate at which mechanical (phonon) energy dissipates via nano- and micro-mechanical support structures plays a critical role in determining the quality of an optomechanical device\cite{cole,bochmann,Rostem2016}. At low temperatures where thermal wavelengths are large compared to the minimal dimension of an elastic waveguide structure, phonon energy transfer is dominated by acoustic phonon modes\cite{rego,schwab}. The physical principles that govern phonon transport in mesoscopic structures are well understood; however, development of full mathematical models for the propagation of phonon modes is only tractable in limited cases\cite{Cross,Judge1,Judge2,wilson-rae}. Even the relatively simple geometry of a narrow rectangular beam joining a wide rectangular cavity cannot be modelled analytically for the full wave vector. The nearest mathematical analysis is that performed by Cross and Lifschitz\cite{Cross}, which evaluates phonon transmission across an abrupt geometric boundary and produces analytic results only for the quasi-static long-wavelength limit. Photiadis and Judge\cite{Judge1, Judge2} expand on this work, analytically evaluating 3-dimensional cantilevered resonators, finding that neglecting thickness does in fact have a significant effect on phonon transmission. This suggests that even for a low-quality factor support structure~\cite{schwab}, assuming 2-dimensional geometries may be an oversimplification. Furthermore, the analysis of Photiadis and Judge\cite{Judge1, Judge2} is only valid for long wavelengths. Thus, a method of accurately analyzing geometrically realistic structures beyond the quasi-static limit remains highly desirable.

 

\begin{figure}
\includegraphics[width = 3.4in]{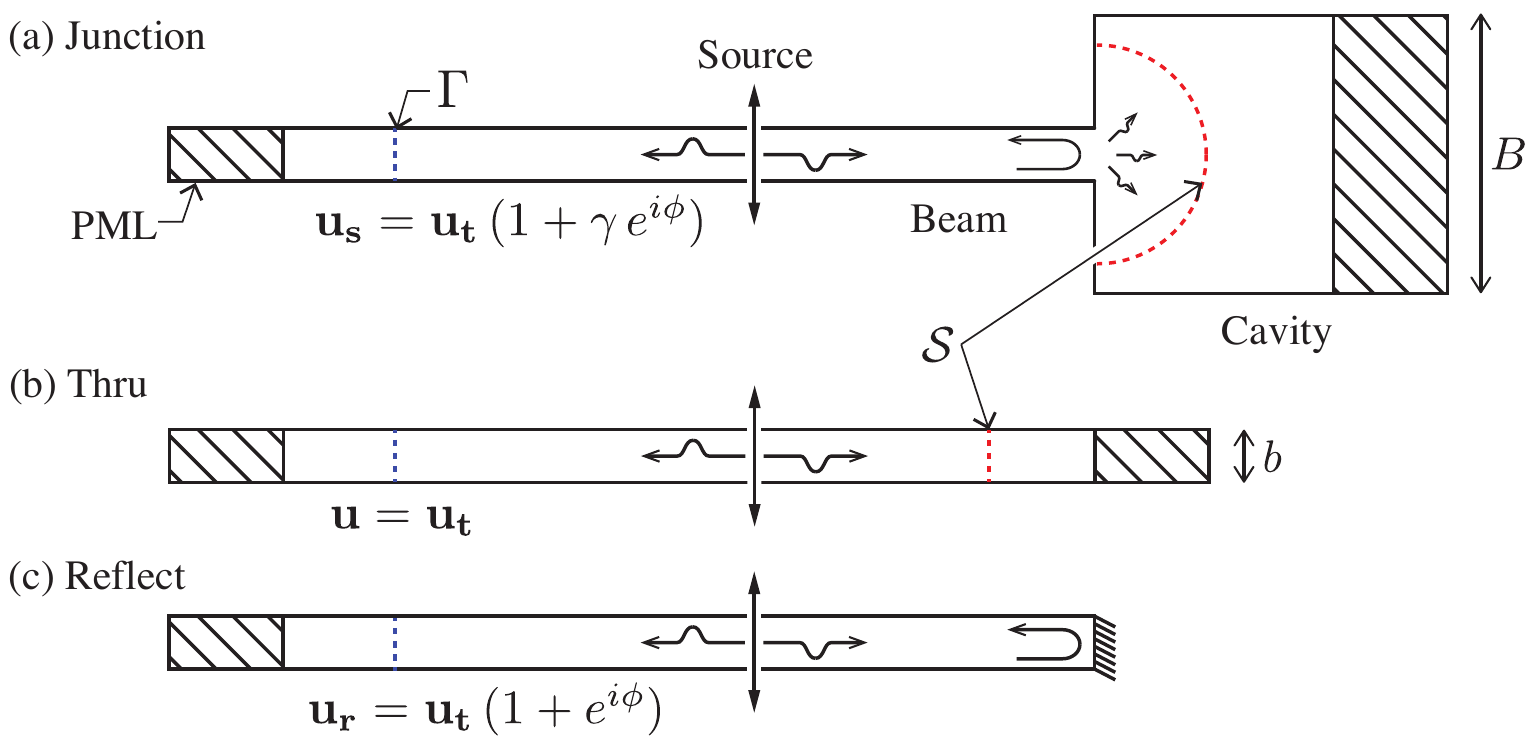}
\caption{\label{fig:reftrans}Plan view of 1-dimensional beams with various terminations. $\Gamma$ and $\mathcal{S}$ are reference planes described in the text. $B/b$ is the ratio of cavity width to beam width. The beam and cavity domains are terminated by a perfectly matched layer (PML).}
\end{figure}

Numerical modeling is one of the most powerful tools available for analysis of phonon transport in structures beyond the simplest cases. This paper explores numerical methods for determining \emph{acoustic} phonon transmission probabilities across an abrupt junction in a rectangular beam (see Fig.~\ref{fig:reftrans}) as a function of frequency using a finite element (FE) approach based on elastic wave theory. The numerical results are compared to the analysis performed by Cross and Lifshitz\cite{Cross}, and explored in terms of phononic heat transfer in a self-suspended mesoscopic device\cite{schwab}. More importantly, the numerical methods established can be readily applied to any geometry and to a wide variety of elastic wave transmission calculations where acoustic modes are scattered into acoustic and optical modes.

\section{Methodology}

The governing equation for elastic wave propagation expressed in the frequency domain is\cite{auld}
\begin{equation}
-\rho\omega^{2}\mathbf{u} = \nabla\cdot\boldsymbol{{\rm T}}+\mathbf{F}
\label{eq:elas}
\end{equation}
where $\rho$ is density, $\omega$ is angular frequency, and $\mathbf{u}$, $\boldsymbol{{\rm T}}$, $\mathbf{F}$ are the displacement vector, stress tensor, and forcing vector. Numerical solutions to Eq. \ref{eq:elas} are readily determined using commercial finite element software. However, using solutions of this formulation to calculate power transmission poses two challenges. First, the phase and amplitude of a propagating displacement wave must be extracted from the complex-valued displacement field solution. The phase-counting method discussed below outlines a procedure for achieving this, while the total power method avoids this issue by using the stress and velocity fields to calculate power transmission directly. Second, in most if not all available finite-element tools, excitation and detection of pure acoustic modes is not possible. Instead, finite geometric loads are typically used, which excite the appropriate acoustic mode, as well as evanescent modes. This implies that phase of the propagating mode cannot simply be assumed to be the product of the propagation constant of the mode and distance away from the source load. Quantification of the error caused by the interaction of reflected waves with the source load is discussed below. 

Finally, we note that relative to time-domain solvers, the frequency-domain methods described here require significantly less computational resources for transmission calculations.

\subsection{Complex Amplitude Reflection (Phase-Counting)}

Figure~\ref{fig:reftrans}a illustrates the reflection of elastic modes from an abrupt junction. To obtain the amplitude of the complex displacement field $\bf u_s$ at the reference plane $\Gamma$, an infinite beam of uniform cross-section (Fig.~\ref{fig:reftrans}b) is modelled and the response $\bf u_t$ is obtained for a given excitation as a function of frequency. $\bf u_t$ provides a normalization for the finite-element model, which is always in the elastic limit but otherwise scaled in response depending on the magnitude of the source load, and the reactance of the elastic waveguide at the excitation frequency.

The displacement field at $\Gamma$ in the presence of the abrupt junction is related to $\bf u_t$,
\begin{equation}
{\bf u_s} = {\bf u_t}\,(1 + \gamma\,e^{i \phi}),
\end{equation}
where $\gamma$ is the amplitude reflection coefficient at the cavity boundary, and $\phi$ is an arbitrary phase offset. To solve for $\gamma$, a second semi-infinite beam with one fixed end is also modeled (Fig.~\ref{fig:reftrans}c), producing the response $\bf u_r$,
\begin{equation}
{\bf u_r} = {\bf u_t}\,(1+ e^{i \phi}).
\end{equation}
The final expression for $\gamma$ is, 
\begin{equation}
{\gamma} = \frac{{\tilde{u_s}}-1}{{\tilde{u_r}}-1},
\label{eq:ref}
\end{equation}
where $\tilde{u_s} = \bf \Vert u_s \Vert/ \Vert u_t \Vert$ and $\tilde{u_r} = \bf \Vert u_r \Vert / \Vert u_t \Vert$ are normalized displacement amplitudes evaluated point-wise at the reference plane $\Gamma$. In the absence of loss, the Fresnel relationship  $\mathcal{T} = 1 - \left|{\gamma}\right|^2$ can be used to obtain the power transmission across the beam-cavity junction.

To quantify the interaction between the source load and reflected fields, a uniform beam with one free end is simulated \cite{Rostem2016,Gaiewski}. The simulation accuracy can be quantified by measuring the deviation of $|\gamma|^2$ from unity. This test provides a robust procedure for refining the finite-element mesh until the desired accuracy is achieved. For the models presented here, an accuracy of $<\,$-30 dB for out-of-plane shear excitation, and less that -50 dB for the other three acoustic modes is readily achieved. This error can be further reduced with the use of a finer mesh, at the expense of increased computation time.

\subsection{Total Power Transmission}
\label{sec:tpt}
An inherent assumption in the phase-counting method is in the symmetry of the elastic scattering region, which excites a single mode both in the beam and in the cavity. As such, the method as illustrated in Fig.~\ref{fig:reftrans} fails for geometries producing multi-moded reflections (e.g. an asymmetric junction). Therefore, a second model is developed to directly evaluate the total transmitted power. The average power (in W\,m$^{-2}$) is given by the Poynting vector\cite{auld}
\begin{equation}
{\bf I} = -\frac{1}{2}Re[i\omega\,\boldsymbol{u}^*\cdot\boldsymbol{{\rm T}}].
\label{eq:pwrflx}
\end{equation}
The total transmitted power is determined by evaluating the following integral over a surface $\mathcal{S}$ (see Fig. \ref{fig:reftrans}) in the cavity enclosing the geometric junction of interest,
\begin{equation}
P = \int_{\mathcal{S}} {\bf I} \cdot {\rm \hat {\bf n}}\,d\mathcal{S},
\label{eq:pwr}
\end{equation}
where ${\rm \hat {\bf n}}$ is the outward-pointing normal vector along $\mathcal{S}$. Power transmission is determined as
\begin{equation}
\mathcal{T} = \frac{P_{\rm J}}{P_{\rm T}},
\label{eq:Tpwr}
\end{equation}
where subscripts J and T refer to the Junction and Thru mechanical power transmission across the reference plane $\mathcal{S}$. This method makes no assumption about the type and number of modes transmitted across the elastic scattering region.

It is important to note that the normalization of the vector field $\bf u$ in Eq.~\ref{eq:ref} and scalar field $P$ in Eq.~\ref{eq:Tpwr} is necessary to obtain accurate scattering coefficients from finite-element computations. At best, neglecting this step may produce severely skewed transmission profiles, and in most cases greater than unity transmission is observed~\cite{yu,yan,lim}. 

\subsection{Model Setup \& Scaling} 

The model of a beam connected to a wider rectangular cavity has received much interest in optomechanical applications~\cite{schwab,Cross,glavin,blanc}. In the example pursued here, the beam is 300 nm wide with a square cross-section. The elastic material properties $E = 280$ GPa, $\rho = 3140$ kg m$^{-3}$ and $\nu = 0.28$ are used to approximate silicon nitride \cite{young}. In developing an FE model, the port geometry and mesh should be scaled to capture the propagating modes with the longest wavelengths of interest. In Fig.~\ref{fig:T}a, the longest wavelength that can be simulated is set by the beam and cavity lengths that are terminated by a perfectly matched layer. Given the dispersion relation for each section, the cavity length is typically much larger than the beam length in a FE model. Here a symmetrically loaded abrupt junction model is considered and only one mode is excited by the beam source and reflected by the abrupt junction. 

Properly establishing a numerical model of the cavity is challenging due to the nontrivial relationship between the cavity modes and excitation frequency of the beam. Moreover, there is the possibility that multiple modes are simultaneously excited within the cavity \cite{Royer2}. 

\begin{figure}
\includegraphics[width = 3.5 in]{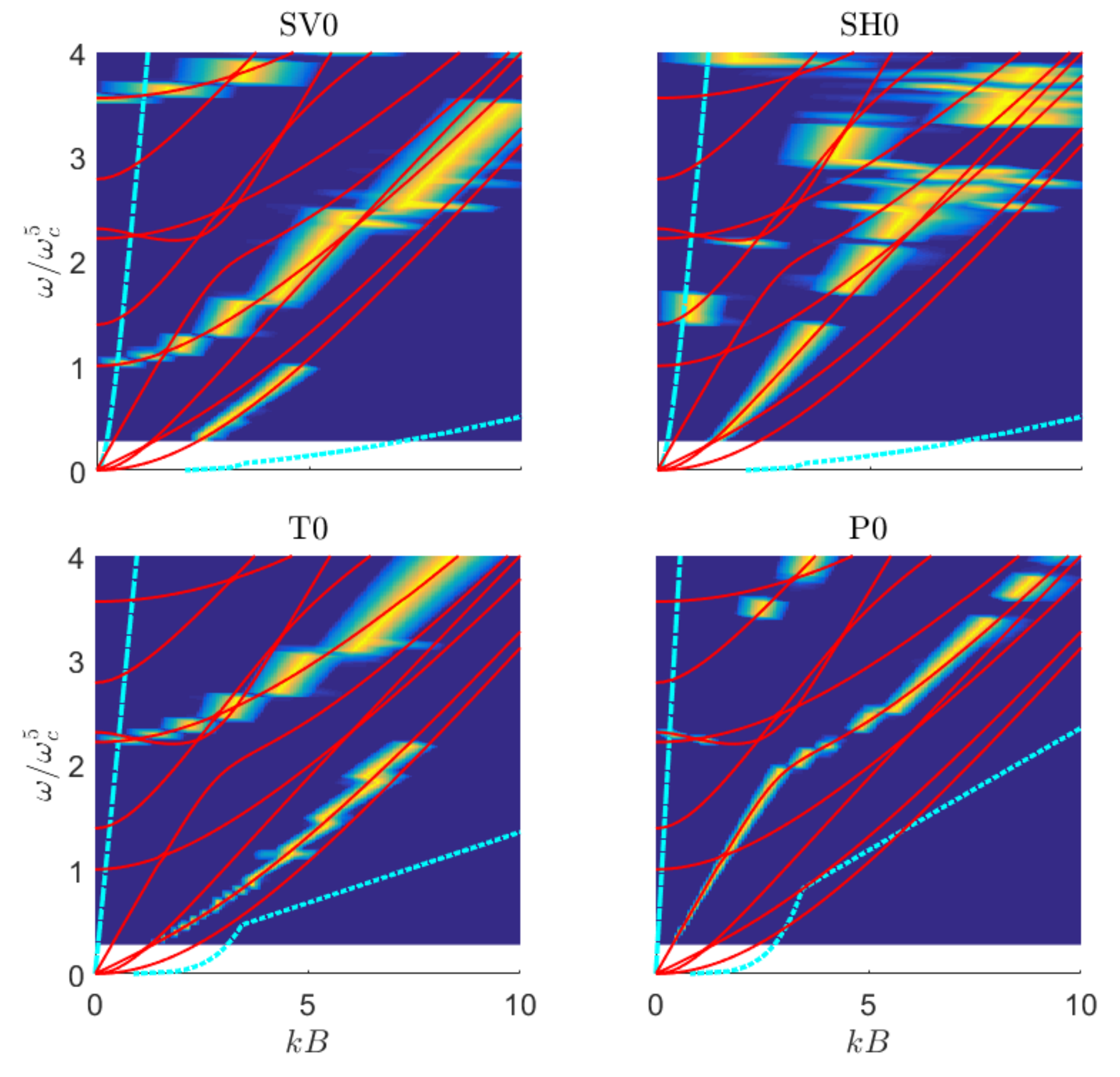}
\caption{\label{fig:Coupling}Excited modes in a cavity for each of the four acoustic modes sourced in the bridge ($B/b=5$). Solid lines show dispersion curves of the cavity. Dotted and dot-dashed lines show meshing and cavity length cutoffs as function of frequency. The discontinuities in the intensity color plot are artifacts of the resolution of the study, which is limited by the number of sampling points used in the Fourier transform of displacement fields. Excited beam excitement are out-of-plane shear (SV0), in-plane shear (SH0), torsional (T0), and compression (P0).}
\end{figure}

To determine which modes are present in the cavity, an in-situ spectroscopic study is performed by exciting the beam at a range of frequencies for each of the four acoustic modes. Figure \ref{fig:Coupling} shows the dispersion relation of the cavity modes and the cutoff imposed by the FE mesh and cavity length. Modes lying between these two limits are effectively captured by the FE model.

For every simulation represented in Fig.~\ref{fig:Coupling}, a Fourier transform of the displacement field is taken along a longitudinal edge of the cavity. The power spectral density output by these transforms are plotted as a function of the beam source frequency. The results clearly show which cavity modes are excited by which beam modes. As is predicted by mode coupling theory\cite{Royer2}, most cases show efficient coupling of a beam mode to a single cavity mode. In general, for any multi-moded elastic waveguiding structure, this type of study is needed to accurately model and capture the response of acoustic and optical modes. 

\section{Results}

\begin{figure}
\includegraphics[width = 3.39 in]{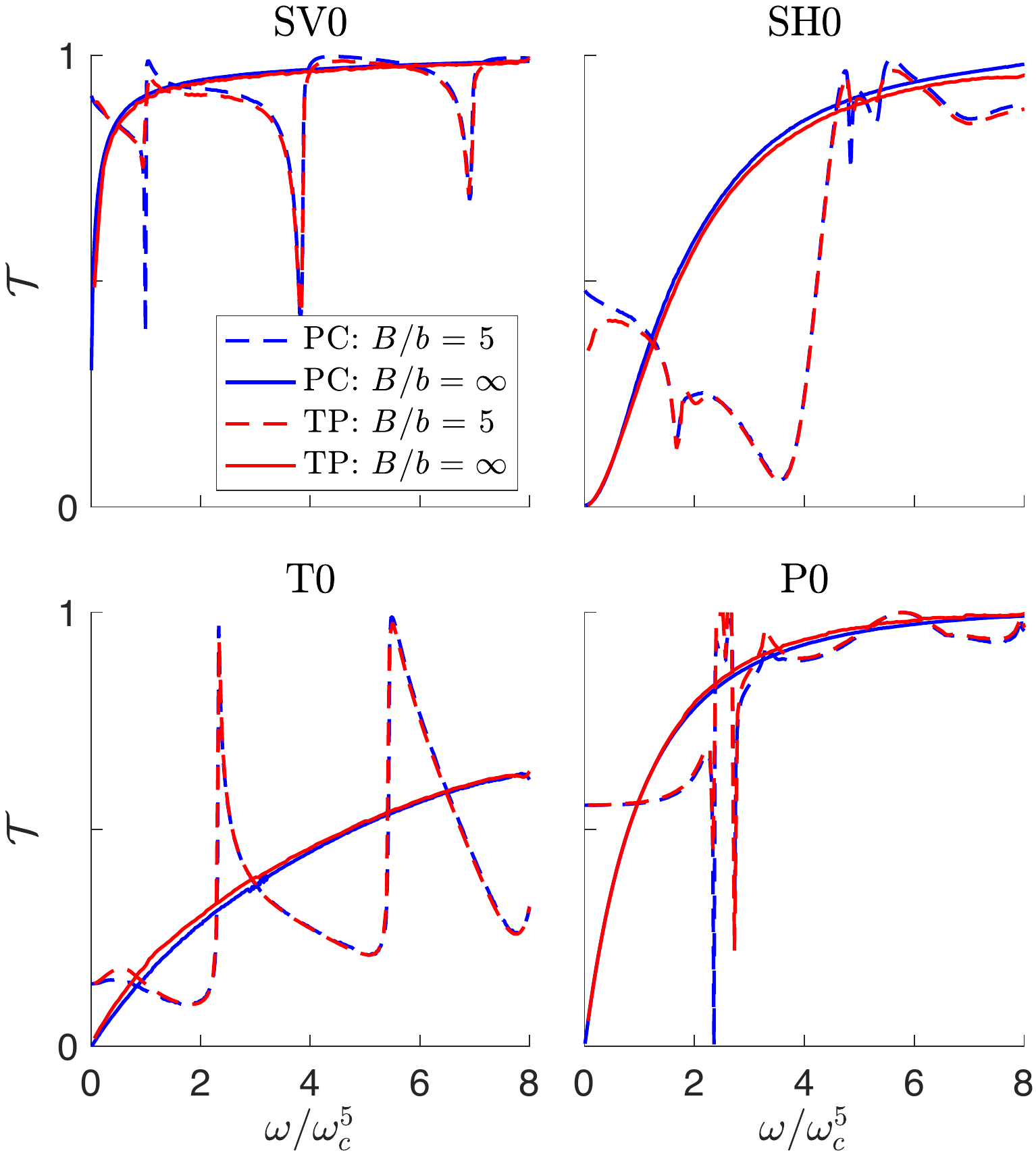}
\caption{\label{fig:T}Power transmission coefficients as a function of frequency. $\omega_c^5$ is the cut-off frequency of the first optical mode in the cavity for $B/b = 5$. Phase-counting (PC) and total power (TP) response are indicted by blue and red lines respectively.}
\end{figure}

\begin{figure}[t]
\includegraphics[width = 3.39 in]{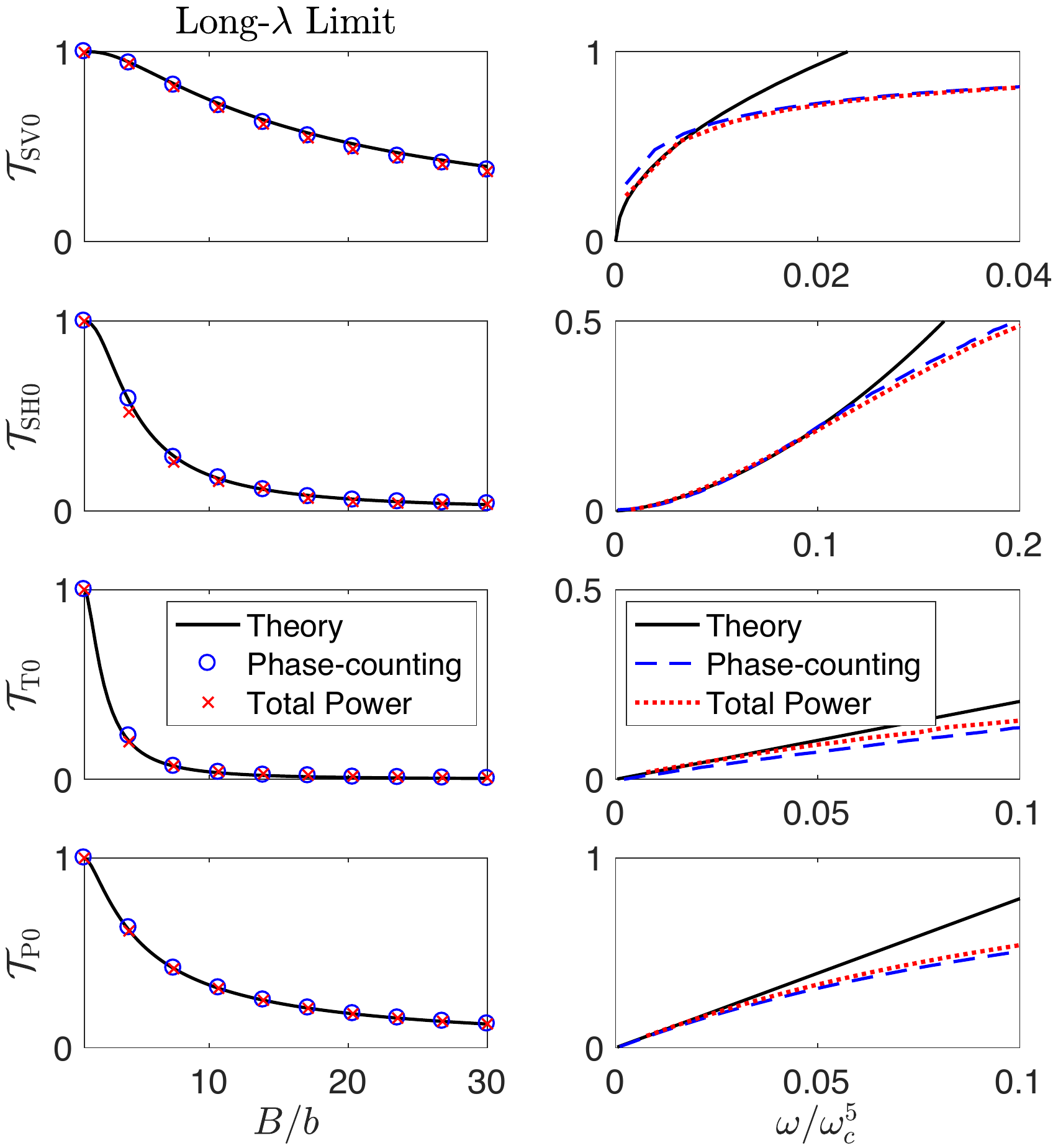}
\caption{\label{fig:theory}Theoretical and simulated transmission coefficients for an abrupt junction. Left column shows relationship as a function of cavity width. Right column shows the relationship for the infinite cavity limit.}
\end{figure}

The power transmission $\mathcal{T}$ of the four acoustic modes across an abrupt junction characterized by $B/b=5$ is shown Fig. \ref{fig:T}. Additionally, transmission into a cavity of infinite width is also shown. All four modes show rejection at low frequencies and rapidly approach unity as frequency is increased and more modes propagate within the cavity. Results of both the phase-counting and total power methods are in good agreement.

Cross and Lifshitz \cite{Cross} derive analytic solutions for the transmission of the four acoustic modes across an abrupt junction using elastic beam theory in the long-wavelength limit. As shown in Fig. \ref{fig:theory}, the theoretical prediction and the results derived from the FE methods described here are in excellent agreement in the long-wavelength limit. In the case of the infinitely wide cavity, the analytic solutions quickly diverge from simulated results as frequency increases, which illustrates the effects of mode coupling to acoustic and optical modes. These effects cannot be captured using elastic beam theory in the long-wavelength limit, where the equations have simple plane wave-like solutions. It is notable that in Fig. \ref{fig:theory}, the total power method shows slightly closer agreement with theoretical predictions than the phase-counting method.

In general, the phase counting method demonstrated quicker convergence than the total power method and produces smoother curves at a given mesh size, as can be seen in Fig. \ref{fig:T}. The main advantage of the total power method lies in its robustness as discussed in Section \ref{sec:tpt}. It also tends to show closer agreement with elastic theory in the long-wavelength limit (see Fig. \ref{fig:theory}).

\section{Thermal Conductance}
\begin{figure}[t]
\includegraphics[width = 3 in]{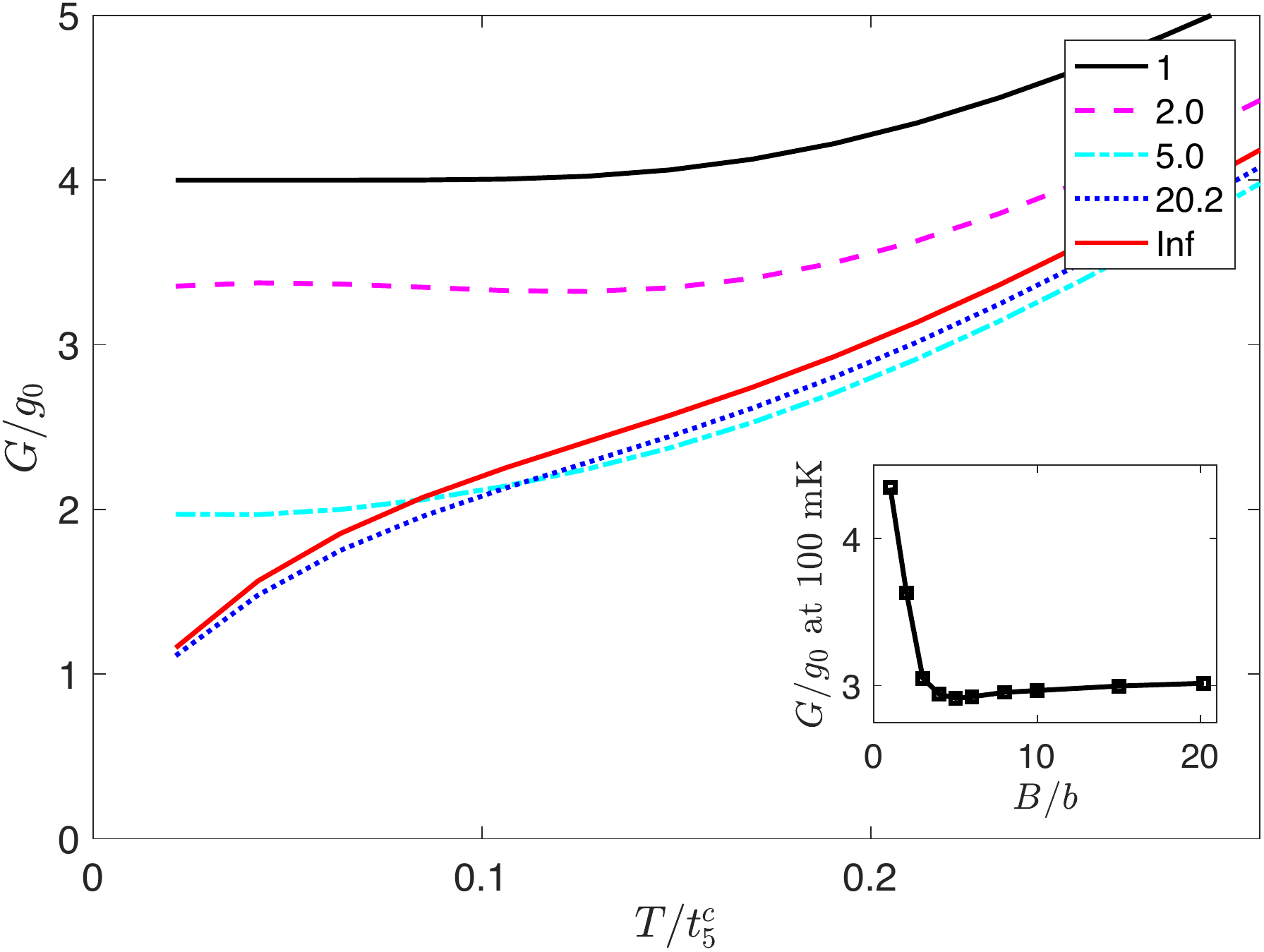}
\caption{\label{fig:cond}Conductance profile of junction for various cavity widths $B/b$, normalized by $g_0 \equiv k_B^2\pi T/(6\hbar)$ and $t_c^5 \equiv \hbar\omega_c^5/k_B$. For the inset, 100 mK is $T/t_c^5 \sim 0.2$.}
\end{figure}

The abrupt junction is of particular interest in the context of heat transfer in mesoscopic structures~\cite{Cross,cole,blanc,puurtinen}. Thermal conductance across a junction can be estimated by summing over the transmission of the propagating modes in the beam:

\begin{equation}
G(T) = \frac{k_{B}^{2}T}{2\pi \hbar} \sum_{\alpha} \int_{x_c^{\alpha}}^{\infty} \mathcal{T}_{\alpha}(x)\,\frac{x^{2}\text{exp}(x)}{[\text{exp}(x)-1]^{2}}\,\text{d}x
\label{eq:cond}
\end{equation}

where $x \equiv \hbar\omega/k_{B}T$, $\omega_c^{\alpha} = x_c^{\alpha}k_{B}T/\hbar$ is the angular cutoff frequency of mode $\alpha$, and $\mathcal{T}_{\alpha}$ is the power transmission through the junction\cite{Rostem2016}. Evaluating this expression produces the relationship between thermal conductance and temperature presented in Fig. \ref{fig:cond} for various values of $B/b$. The conductance of a uniform beam ($B/b = 1$) is included as a point of reference, and at very low temperatures, is defined by the quantum of conductance~\cite{schwab}. In evaluating Equation \ref{eq:cond} we compute transmission for $\alpha = [1,4]$ up to $\omega_c^5$ in the beam. Past $\omega_{c}^{5}$, transmission is assumed to be unity. Optical modes of the beam up to 20 GHz are included in all calculations. As $B/b$ is increased, the thermal conductance quickly converges to the value obtained for a cavity of infinite width. A notable result from the inset in Fig. \ref{fig:cond} is that the minimum in conductance in the 0.05 -- 0.15 K range occurs at $B/b \approx 5$ due to the strong rejection of low-frequency waves at this cavity width, particularly for the in-plane shear mode.

\section{Conclusions}
Two methods have been explored for obtaining robust simulations of acoustic phonon transmission in elastic waveguiding structures using FE computations. The accuracy of each method is evaluated using the model of an abrupt geometric junction. In the long-wavelength limit, the behavior of the power transmission is in close agreement with the available analytic results for a square beam connected to a wider cavity. The two methods can be successfully applied to compute the transmission profiles of acoustic modes in a beam at frequencies below the first optical mode. More generally and in the context of high sensitivity applications, the numerical methods described here provide versatile and reliable tools in the analysis of heat transport in phononic crystal structures, and energy storage in mesoscopic resonators. 


\begin{acknowledgments}
We gratefully acknowledge financial support from the NASA Astrophysics Research and Analysis (NNX17AH83G) grant. We thank Tuomas Puurtinen for useful discussions.
\end{acknowledgments}

\nocite{*}
\bibliography{Junction}

\end{document}